# Weakly Coupled Type-II Superconductivity in a Laves compound ZrRe$_2$

Yingpeng Yu[1,2*], Zhaolong Liu[1,3], Qi Li[1,3], Zhaoxu Chen[1,4], Yulong Wang[1,4], Munan Hao[1,3], Yaling Yang[1,4], Chunsheng Gong[5], Long Chen[1,2], Zhenkai Xie[1,2], Kaiyao Zhou[1,2], Huifen Ren[1,2], Xu Chen[1,2], and Shifeng Jin[1]

[1]*Beijing National Laboratory for Condensed Matter Physics and Institute of Physics, Chinese Academy of Sciences, Beijing 100190, China*
[2]*University of Chinese Academy of Sciences, Beijing 100049, China*
[3]*College of Materials Science and Opto-Electronic Technology, University of Chinese Academy of Sciences, Beijing 101408, China*
[4]*School of Physical Sciences, University of Chinese Academy of Sciences, Beijing 100049, China*
[5]*Beijing Lattice Semiconductor Co., LTD*

We present a comprehensive investigation of the superconducting properties of ZrRe$_2$, a Re-based hexagonal Laves compounds. ZrRe$_2$ crystallizes in a C14-type structure (space group P6$_3$/mmc), with cell parameters a=b=5.2682(5) Å and c=8.63045 Å. Resistivity and magnetic susceptibility data both suggest that ZrRe$_2$ exhibits a sharp superconducting transition above 6.1 K. The measured lower and upper critical fields are 6.27 mT and 12.77 T, respectively, with a large upper critical field that approached the Pauli limit. Measurements of the heat capacity confirm the presence of bulk superconductivity, with a normalized specific heat change of $\Delta C_e/\gamma T_c = 1.24$ and an electron-phonon strength of $\lambda_{ep} = 0.69$. DFT calculations revealed that the band structure of ZrRe$_2$ is intricate and without van-Hove singularity. The observed large specific heat jump, combined with the electron-phonon strength $\lambda_{ep}$, suggests that ZrRe$_2$ is a weakly coupled type II superconductor.

# 1. Introduction

Within the realm of intermetallic compounds, there exists a significant family of compounds known as Laves phases, which share the general formula $AB_2$. It is reported that more than 300 kinds of Laves phases have been found in the binary alloys $AB_2$, and more than 900 kinds in the ternary alloys $A(B,C)_2$, considering the third alloying elements dissolve into the matrix. From the viewpoint of crystalline structure, only three types of Laves structures have been frequently observed to date: hexagonal C14-type ($MgZn_2$-type), cubic C15-type ($MgCu_2$-type), and hexagonal C36-type ($MgNi_2$-type), as shown in Fig. 1. The versatility in crystal structure and composition in different laves phases has resulted in their extensive utilization in various fields, such as magnetic materials, hydrogen storage materials, and superconducting magnets etc. [1-9].

Given the thousands of materials that are crystallized in Laves phases, exploring superconductivity in these phases presents an intriguing challenge and has the potential to offer valuable insights into the unique superconducting phenomena observed in some Laves phase materials. While previous research studies have predominantly focused on different C15-type materials [10-17], increasing studies have been conducted on materials based on the C14 type [18-22]. Within the C14-type Laves phases, several Ru-based superconductors have been identified with effective electron correlation and a maximum $T_c$ of 2.4 K [20]. More recently, a series of Os-based superconductors with a maximum $T_c$ of 2.90 K has also been uncovered [22]. In addition, there are also some reports on the superconductivity in the ternary C14 Laves phase, such as $Ta_2V_{3.1}Si_{0.9}$ [23], and several ordered Laves phases, for example $Mg_2Ir_3Si$ [24] and $Li_2IrSi_3$ [25], along with some partially ordered phase, including $Mg_2Ir_{2.3}Ge_{1.7}$ [26].

Recently, unconventional time-reversal symmetry-breaking superconductivity has been proposed to exist in the frustrated C14-type superconductor, $HfRe_2$ [18], sparking further investigations into superconductors in Re-based Laves metals to explore unconventional behavior. Additionally, there are other Re-based compounds, such as $ZrRe_2$, known to exhibit superconductivity at relatively high temperatures. While its critical temperature ($T_c$) and lattice parameters have been reported [27, 28], detailed information on its superconducting properties remains undisclosed. This underscores the importance of conducting in-depth research to uncover the detailed superconducting characteristics of $ZrRe_2$.

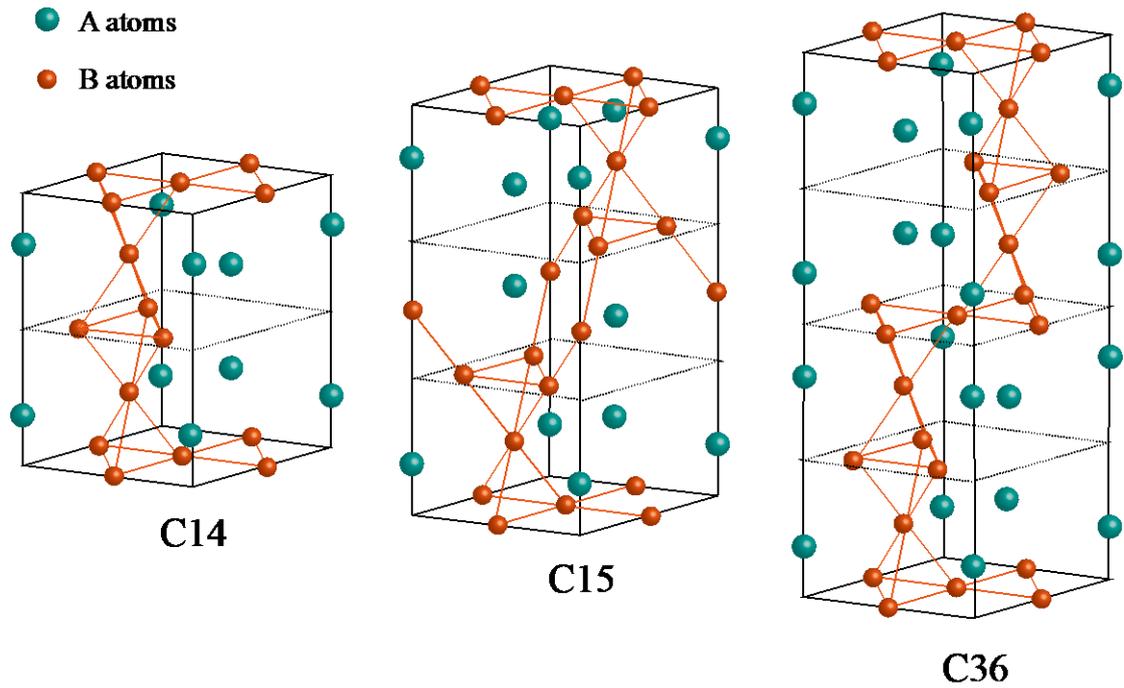

Fig. 1. (Color online) Three distinct polytypes of the Laves phase structure arranged in a hexagonal configuration. The large circles representing A atoms and the small circles representing B atoms are interconnected.

Based on the above, we present a comprehensive analysis of the superconducting properties of ZrRe$_2$ in this paper. Through a detailed examination of magnetism, resistivity, and specific heat, we confirm a weakly-coupled type-II superconductivity occurred below the transition temperature of 6.1 K. Notably, the unusually large upper critical field values near the Pauli limit suggest the potential presence of unconventional behavior, akin to the observations in HfRe$_2$ [18] and distinct from most other C14-type superconductors [19-22]. Furthermore, DFT calculations have excluded the relevance of VHS band or flat band to the emergence of high superconductivity in this C14-type Laves phase metal.

## 2. Experimental/Methods

We synthesized polycrystalline ZrRe$_2$ samples using an arc melting method. Zirconium (99.9% purity) and rhenium (99.9% purity) were thoroughly mixed in a 1:2 molar ratio and pressed into pellets. These pellets underwent at least six arc-melting cycles in a high-purity argon atmosphere with periodic turnovers, resulting in weight losses of less than 1%. Unlike other samples, ours did not require annealing. The final products exhibited a metallic luster and remained stable in ambient air.

Powder X-ray diffraction (pXRD) data was collected at room temperature using a PANalytical X-ray diffractometer with Cu-Kα radiation. The crystal structure was refined using the open-source software package FULLPROF [29], and crystal structure diagrams were generated using VESTA [30]. Measurements of transport, magnetic, and thermodynamic properties were conducted using a Physical Property Measurement System (PPMS, Quantum Design) and a Vibrating Sample Magnetometer (VSM, Quantum Design). It is important to note that all data presented in this paper were obtained from samples within the same batch.

First principles calculations were carried out using density functional theory (DFT) implemented in the Vienna ab initio simulation package (VASP) [31]. We adopted the generalized gradient approximation (GGA) in the form of the Perdew-Burke-Ernzerhof (PBE) [32] for the exchange correlation potentials. The projector-augmented-wave (PAW) [33] pseudopotentials were used with a plane wave energy of 600 eV. The $4d^2 5s^2$ electronic configuration of Zr and $5d^5 6s^2$ electronic configuration of Re were treated as valence electrons, respectively. A Monkhorst-Pack Brillouin zone sampling grid [34] with a resolution of $0.02 \times 2\pi$ Å$^{-1}$ was applied. Atomic positions and lattice parameters were relaxed until all the forces on the ions were less than $10^{-2}$ eV/Å. In calculating the electronic structure, the spin-orbit coupling (SOC) effect was included.

## 3. Results
### 3.1. Structural characterization

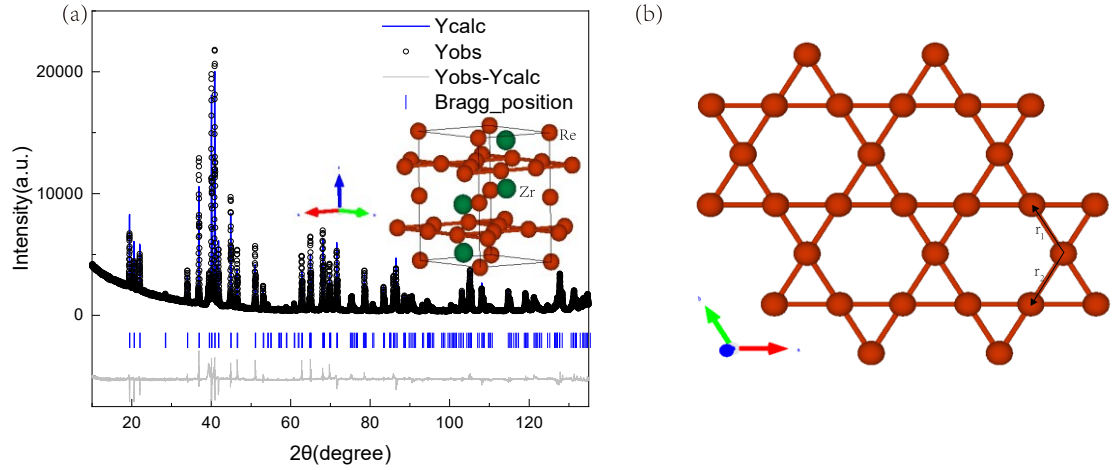

Fig. 2. (a) Room-temperature pXRD pattern of ZrRe$_2$ and its Rietveld refinements. The inset displays the conventional unit cell. (b) Re atoms form a special lattice, in which r$_1$ ≠ r$_2$.

Fig. 2. (a) illustrates the pXRD pattern for ZrRe$_2$. The pattern can be effectively refined to match a C14-type Laves phase structure (MgZn$_2$-type, space group P6$_3$/mmc), confirming the successful formation of the intended ZrRe$_2$ phase. A Rietveld refinement process produced cell parameters $a = b = 5.2682(5)$ Å, $c = 8.6304(5)$ Å, which closely match an earlier experimental finding [28]. Additional information on the results of refinement can be found in Table 1. The refined crystal structure of ZrRe$_2$ is shown as the inset of Fig. 2(a). The structure of ZrRe$_2$ consists of Zr$_2$Re layers and Re atoms layers stacking along the crystallographic c axis, as shown in Fig. 2(a). We can see that two distinct Re sites exist within the structure in ZrRe$_2$. Notice that one of the Re site form a special lattice, which is clearly shown in Fig. 2(b). There are two distinct sets of Re–Re bond lengths within this Re atoms layer, in which the shorter and longer distances of the Re-Re bonds are 2.58 Å and 2.68 Å, respectively, and obviously the difference between the two kinds of Re-Re bonds is only 3.7%.

Table 1. Crystallographic parameters of ZrRe$_2$ from Rietveld refinement of pXRD.

| Chemical formula | ZrRe$_2$ |
|---|---|
| Formula weight | 463.64 |
| Z | 4 |
| R$_{wp}$ | 7.23% |
| Space group | P6$_3$/mmc (No. 194) |
| a (experimental)(Å) | 5.2682(5) |
| c (experimental)(Å) | 8.6304(5) |
| V (Å$^3$) | 207.3 |

*3.2 Superconducting properties*

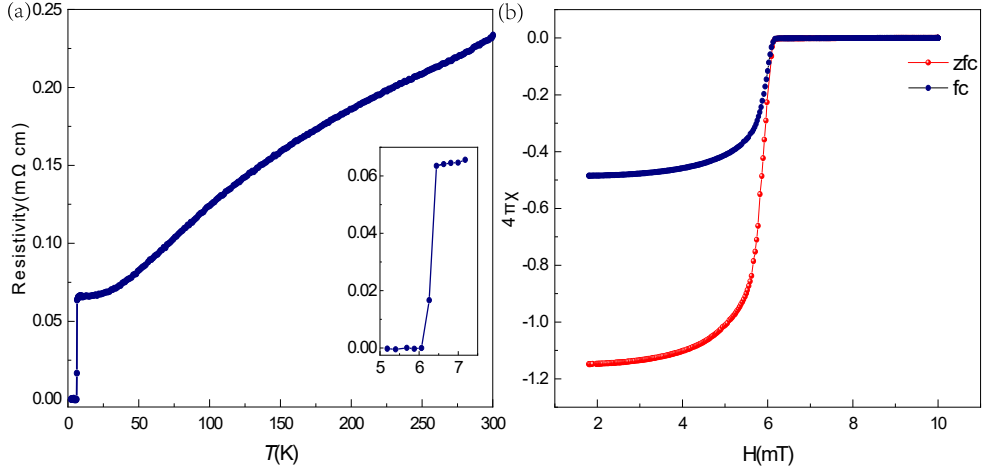

Fig. 3. (a) Temperature dependence of electrical resistivity $\rho(T)$ at zero field for ZrRe$_2$. Insets: enlarged view of $\rho(T)$ curves between 4.0 K and 8.0 K. (b) Temperature dependence of magnetic susceptibility $4\pi\chi$ for ZrRe$_2$ measured at 1 mT with zero-field-cooling (ZFC) and field-cooling (FC) modes.

Fig. 3(a) illustrates the temperature-dependent resistivity $\rho(T)$ of ZrRe$_2$ across the range of 1.8 K to 300 K. The observed monotonic decrease in resistivity towards lower temperatures suggests a metallic behavior, while the relatively small residual resistivity ratio (RRR) of 3.53 (calculated as 2.33/0.66) is attributed to grain boundary effects. At elevated temperatures, the $\rho(T)$ curves exhibit a saturation tendency, possibly linked to the Ioffe-Regel limit [35]. This limit implies that the charge carrier mean free path is comparable to the interatomic spacing [36]. A notable resistivity drop occurs at 6.44 K (see inset in Fig. 3(a)), indicating the onset of a superconducting transition. Additionally, magnetic susceptibility $4\pi\chi$ curves in Fig. 3(b) reveal diamagnetic transitions at 1 mT, with the determined $T_{c,onset}$ at 6.16 K, consistent with values obtained from the $\rho(T)$ curve and the heat capacity measurements shown below. A complete Meissner state is achieved at around 4.5 K. After considering the demagnetization factor, the magnetic field inside the superconductor can be modified according to the formula: $H_i = H_a(1-n)$, where $H_i$ is the magnetic field inside the superconductor and $H_a$ is the magnetic field outside the superconductor. And n is the demagnetization factor, which we take as $\frac{1}{3}$. And thus the zero-field-cooling (ZFC) curve at 2.0 K indicates a superconducting volume fraction (SVF) of approximately 115%, suggesting bulk superconductivity. The large SVF observed in field-cooling (FC) mode implies weak flux pinning effects.

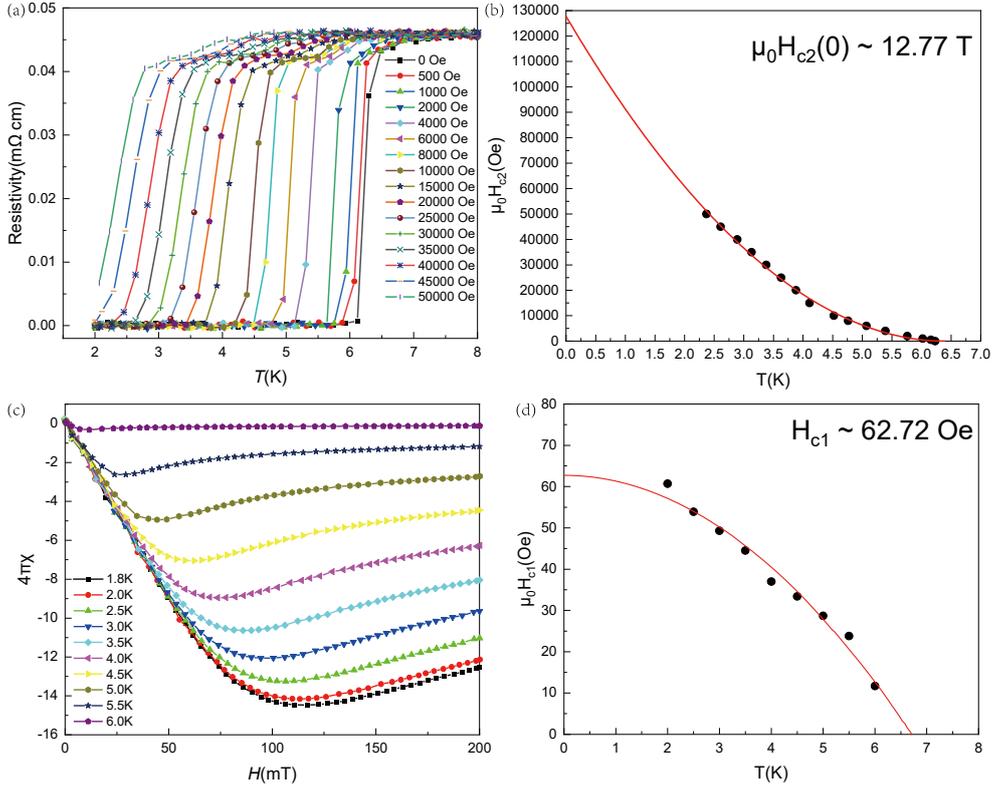

FIG. 4. (a) Temperature dependence of the resistivity at a various external field. (b) Transition temperatures extracted from (a) and the respective fitting by using the empirical formula[37]. (c) Magnetic hysteresis of ZrRe$_2$ at various temperature. (d) Lower critical field $\mu_0 H_{c1}$ of ZrRe$_2$ as a function of temperature. The red line is a linear fit.

Fig. 4(a) illustrates the decrease in critical temperature ($T_c$) as the external magnetic field is raised from 0 to 5 T. We used the $T_c^{mid}$ as the transition temperature and summarized them into Fig. 4(b). Intriguingly, H$_{c2}$ of ZrRe$_2$ has an upward curvature, which has also been observed in many superconductors like Bi$_5$O$_4$S$_3$Cl [38], EuFBiS$_2$ [39], Bi$_4$O$_4$S$_3$ [40, 41], La$_{1-x}$M$_x$OBiS$_2$(M = Th, Hf) [42], HfRe$_2$ [18], MgB$_2$ [43], and Td-MoTe$_2$ [44]. In MgB$_2$ [43], Td-MoTe$_2$ [44], and HfRe$_2$ [18], this upward curvature of H$_{c2}$ is considered to be a signature of multi-band superconductivity, implying that an unconventional pairing mechanism may exist in Re-based superconductors. Deviating from the Werthamer–

Helfand–Hohenberg theory based on the single-band model [45], the $H_{c2}$ curve of ZrRe$_2$ can be well fitted by using the empirical expression[37]:

$$\mu_0 H_{c2} = \mu_0 H_{c2}(0)(1-t)^{1+\alpha}. \tag{1}$$

$t = \frac{T}{T_c}$, which gave an upper critical field of 12.77 T. The Pauli-Clogston field is:

$$\mu_0 H_p(0) = 1.84 T_c = 11.8\ T. \tag{2}$$

Which is nearly equal to the field we obtain from the empirical formula. The coherence length ($\xi$) can be acquired from Ginzburg-Landau (GL) equation :

$$\mu_0 H_{c2}(0) = \frac{\Phi_0}{2\pi \xi_0^2}. \tag{3}$$

Where $\Phi_0 = 2.07 \times 10^{-15}\ T\ m^2$, is the magnetic flux unit. By using $\mu_0 H_{c2}(0) = 12.77$ T, the calculated $\xi$ is 5.08 nm. The magnetization vs external field over a range of temperatures below $T_c$ is presented in Fig. 4(c). The field deviates from a linear curve of full Meissner effect were deemed as the lower critical field at each temperature and was summarized in Fig. 4(d). The temperature-dependent lower critical field is well fitted with the Ginzburg-Landau equation described as follows:

$$\mu_0 H_{c1} = \mu_0 H_{c1}(0)\left[1 - \left(\frac{T}{T_c}\right)^2\right] \tag{4}$$

Which gives $\mu_0 H_{c1}(0) = 62.72$ Oe. By using formula

$$\mu_0 H_{c1}(0) = \frac{\Phi_0}{4\pi \lambda^2} \ln\left(\frac{\lambda}{\xi}\right). \tag{5}$$

we get the penetration depth $\lambda = 331\ nm$. The calculated GL parameter of $\kappa = \frac{\lambda}{\xi} \sim 61.61$ confirms the type-II superconductivity in ZrRe$_2$.

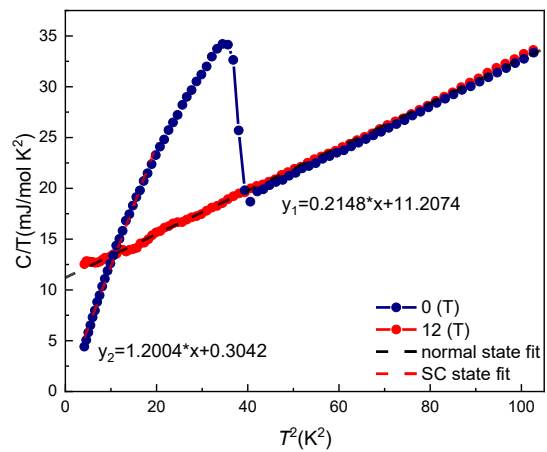

Fig. 5. Heat capacity of ZrRe$_2$ from 2.0 to 10 K. The red line is measured at 12 T.

Fig. 5 presents the analysis of the superconductivity transition through heat-capacity measurements. The presence of a clear anomaly at 6.27 K signifies the bulk superconductivity. We applied a high magnetic field of $\mu_0 H = 12\,T$ to suppress the superconductivity and fitted the normal-state heat capacity according to

$$\frac{C_p(T)}{T} = \gamma + \beta T^2. \tag{6}$$

The extrapolation gave $\gamma = 11.2074\,mJ/(mol\,K^2)$ and $\beta = 0.2148\,mJ/(mol\,K^4)$, where the $\gamma$ is Sommerfeld electronic specific heat coefficient and $\beta$ is the phonon specific heat coefficient. We also fitted the superconducting state and obtained the fitting parameters. Together with the fitting parameters of the normal state, we can obtain the superconducting volume fraction under the specific heat test results: 97.28(5)%. This proves the bulk superconducting properties of ZrRe$_2$ again. By using the formula

$$\beta = N \frac{12\pi^4}{5\Theta_D^3} R. \tag{7}$$

and N = 3 for ZrRe$_2$, we got a Debye temperature of $\Theta = 301\,K$. Notably, the specific heat jump of $\Delta C_e/\gamma T_c = 1.24$ is lower than that of the BCS anticipated value of 1.43, indicating a weak electron-phonon coupling. We can further estimate the electron–phonon coupling constant $\lambda_{ep}$ using the McMillan relation [46]

$$\lambda_{ep} = \frac{1.04 + \mu^* \ln\left(\frac{\Theta_D}{1.45 T_c}\right)}{(1 - 0.62\mu^*) \ln\left(\frac{\Theta_D}{1.45 T_c}\right) - 1.04}. \tag{8}$$

where $\mu^*$ is the Coulomb screening parameter (we here set it to 0.13). $\lambda_{ep} = 0.69$ is thus obtained, indicating that ZrRe$_2$ hosts a weak to moderate coupling strength. Moreover, the DOS at Fermi level (E$_F$) is estimated using $N(E_F) = 3\gamma/[\pi^2 k_B{}^2 (1 + \lambda_{ep})]$ based on $\gamma$ and $\lambda_{ep}$, yielding $N(E_F) = 2.8114\,eV^{-1} \cdot f.u.^{-1}$. Table 2 summarizes the superconducting parameters of ZrRe$_2$ and other superconducting materials in the AB$_2$ type Laves phase. From the table, it can be observed that ZrRe$_2$ exhibits the highest T$_c$, the largest upper critical field and Ginzburg-Landau parameter (GL), and the highest Debye temperature.

Table 2. Superconducting and thermodynamic parameters of ZrRe$_2$ compared with other AB$_2$ Laves compounds.

| Parameter | Units | ZrRe$_2$ | LuRu$_2$ | LuOs$_2$ | BaRh$_2$ | BaIr$_2$ | ThIr$_2$ |
|---|---|---|---|---|---|---|---|
| $T_c$ | K | 6.44 | 2.23 | 3.47 | 5.6 | 2.7 | 5.58 |
| $\mu_0 H_{c1}(0)$ | mT | 6.27 | 8.0 | 4.8 | 5.3 | 28.6 | 9.5 |
| $\mu_0 H_{c2}(0)$ | T | 12.77 | 2.31 | 1.64 | 4.63 | 6.77 | 2.25 |
| $\xi_{GL}$ | Å | 50.8 | 387 | 141 | 84.3 | 69.7 | 12.1 |
| $\lambda_{GL}$ | Å | 3313 | 2870 | 3270 | 3400 | 1520 | 2470 |
| $\kappa_{GL}$ |  | 61.61 | 7.6 | 23 | 40.3 | 21.8 | 20.4 |
| $\gamma$ | $mJ/(mol\ K^2)$ | 11.2 | 17.5 | 14.5 | 25.7 | 12.0 | 13.1 |
| $\Theta_D$ | K | 301 | 250 | 270 | 178 | 147 | 230 |
| $\lambda_{ep}$ |  | 0.69 | 0.57 | 0.59 | 0.801 | 0.63 | 0.72 |
| $\Delta C_{es}/\gamma T_c$ |  | 1.24 | 1.26 | 1.41 | 1.86 | 1.2 | 1.19 |
| $N(E_F)$ | $states\ eV^{-1}\ f.u.^{-1}$ | 2.81 | 3.73 | 3.86 | 6.1 | 3.12 | 3.23 |

*3.3 First-principles calculations*

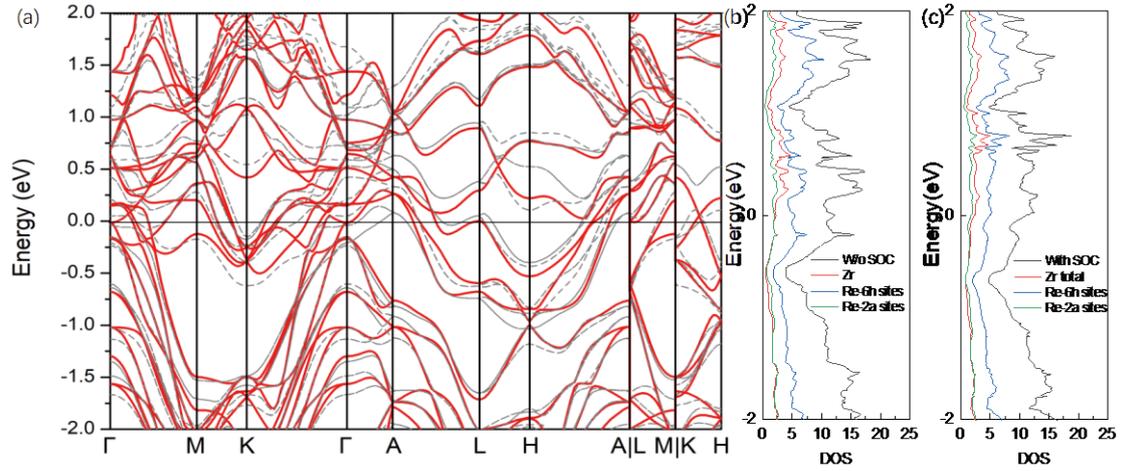

Fig. 6. (a) Calculated electronic band structure of ZrRe$_2$ without and with SOC near the Fermi level. The corresponding DOS plots are shown in (b) and (c), respectively.

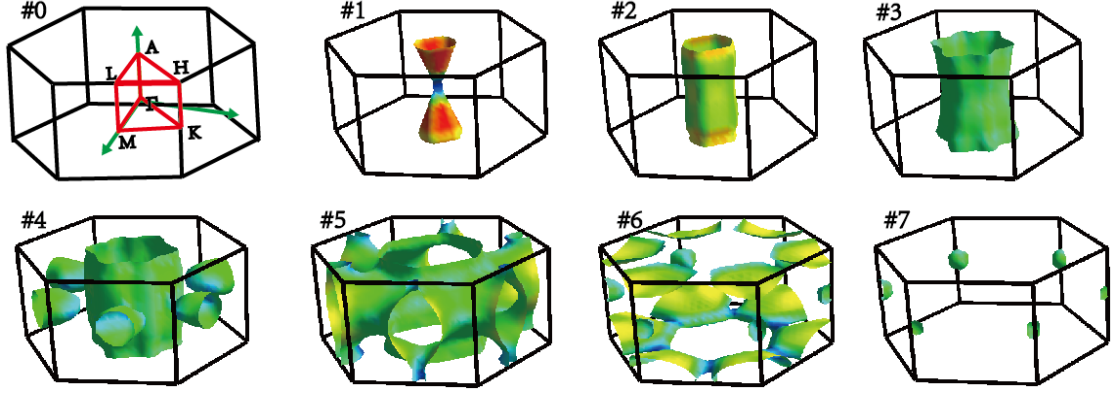

Fig. 7. Theoretical results of the FS topology: computed FS topologies for different bands in the first Brillouin zones for calculations without SOC.

To better understand the properties of ZrRe$_2$, we performed the DFT calculations for the electronic band structures with and without the spin orbit coupling (SOC) effects. The band structures of ZrRe$_2$ near the Fermi level ($E_F$) are depicted in Fig. 6(a). The results without spin-orbit coupling (SOC) are represented by dashed lines, while the solid lines depict the results with SOC. It's noteworthy that the inclusion of SOC brings about significant changes in the band dispersion near $E_F$. Fig. 6(b) and (c) illustrate the calculated total density of states (DOS) for ZrRe$_2$. Similar DOS patterns were observed in other AB$_2$ structure C14 type compounds, such as ZrOs$_2$ [22], LuOs$_2$ [21], YRu$_2$ [20], indicating their analogous crystal structures and electric structures. The DOS near the Fermi energy in ZrRe$_2$ primarily originates from the Re 5d states, particularly the Re 5d states in 6h sites, with Zr states making relatively minor contributions. It is worth noting that, from our DFT calculation results, whether considering the spin orbit coupling (SOC) or not, we are unable to observe the Van Hove singularity band near the Fermi surface, nor can we observe flat bands or the presence of Dirac cones. Therefore, regarding the speculation on the possible correlation between high $T_c$ and the existence of Van Hove singularities near the Fermi surface, in the case of ZrRe$_2$ this speculation can be safely refuted. Moreover, our DFT calculations reveal the presence of 4 Fermi surface sheets for the SOC split bands and 7 Fermi surfaces without SOC. As shown in Fig. 7, all seven Fermi surfaces exhibit substantial three-dimensional behavior, with FS 4 and FS 5 having the largest area among all the FS, similar to the findings in HfRe$_2$. Considering the energy bands from Re dominate the DOS of ZrRe$_2$ around Fermi energy, the superconducting states in ZrRe$_2$ are predominantly attributed to the Re atom states.

4. Conclusion

In this work, we synthesize and analyze the superconducting properties of ZrRe$_2$. The compound exhibits a sharp superconducting transition above 6.1 K, with lower and upper critical fields of 6.27 mT and 12.77 T, respectively. The unusually large upper critical field values near the Pauli limit suggest the potential presence of unconventional behavior, similar with observations in HfRe$_2$ and distinct from most other C14-type superconductors. The measured heat capacity confirms the presence of bulk superconductivity, with a normalized specific heat change of $\Delta C_e/\gamma T_c = 1.24$ and an electron-phonon strength of $\lambda_{ep} = 0.69$. Furthermore, our DFT calculations have excluded the relevance of van-Hove singularity or flat band to the emergence of high superconductivity in this C14-type Laves phase metal.

**Acknowledgements**

This work is supported by the National Natural Science Foundation of China (Grant No. 52272268), the Strategic Priority Research Program and Key Research Program of Frontier Sciences of the Chinese Academy of Sciences (Grant No. XDB33010100), the Informatization Plan of Chinese Academy of Sciences (Grant No. CAS-WX2021SF-0102), and the Synergetic Extreme Condition User Facility (SECUF).

*E-mail: yingpengyu93@gmail.com